\documentclass[11pt,twoside]{article}

\usepackage{asp2006}
\usepackage{epsf}
\usepackage{psfig}
\usepackage{lscape}

\begin{document}
\title{Balmer Absorption Lines in FeLoBALs}
\author{K.~Aoki\altaffilmark{1}, I.~Iwata\altaffilmark{2}, K.~Ohta\altaffilmark{3}, N.~Tamura\altaffilmark{1}, M.~Ando\altaffilmark{3}, M.~Akiyama\altaffilmark{1}, G.~Kiuchi\altaffilmark{3}, and K.~Nakanishi\altaffilmark{4}}
\altaffiltext{1}{Subaru Telescope, National Astronomical Observatory of Japan (NAOJ), 650 N. A'ohoku Place, Hilo, HI, 96720, U.S.A.}
\altaffiltext{2}{Okayama Astrophysical Observatory, NAOJ, Okayama, 719-0232, Japan}
\altaffiltext{3}{Department of Astronomy, Kyoto University, Kyoto, 606-8502, Japan}
\altaffiltext{4}{Nobeyama Radio Observatory, NAOJ, Minamimaki Minamisaku, Nagano, 384-1305, Japan}

\begin{abstract} 
We discovered non-stellar Balmer absorption lines in two 
many-narrow-trough FeLoBALs (mntBALs) 
by the near-infrared spectroscopy with Subaru/CISCO. 
Presence of the non-stellar Balmer absorption lines is known to date only in the Seyfert galaxy NGC 4151, 
thus our discovery is the first cases for quasars. 
Since all known active galactic nuclei with Balmer absorption lines share 
characteristics, it is suggested that there is a population of BAL quasars which have unique structures at their nuclei 
or unique evolutionary phase.
\end{abstract}


\section{Broad Absorption Line Quasar}
Broad absorption line (BAL) quasars are quasars which show absorption lines
in UV resonance lines, with velocity widths of $\sim 2000 - 20000$ km s$^{-1}$.
BAL quasars are divided into three subtypes.
High-ionization BAL quasars show absorption from C {\small IV}, N {\small V}, Si {\small IV} and Ly $\alpha$.
Low-ionization BAL quasars (LoBALs) show absorption from Mg {\small II}, Al {\small III}, and Al {\small II},
in addition to the above high-ionization absorption.
A fraction of LoBALs shows absorption from Fe {\small II} and Fe {\small III}.
They are called iron LoBALs (FeLoBALs).
\par 
\citet{Hal02} discovered unique FeLoBALs which have tremendous number of 
absorption lines, so-called ``many-narrow-trough FeLoBAL'' (mntBAL). 
We have searched for mntBALs by visual inspection of $\sim 20000$ spectra of
quasars in $2.1 \leq z \leq 2.8$ and $1.3 \leq z \leq 1.65$ in the SDSS
database.
H$\beta$ and H$\alpha$ emission lines in these redshifts are expected to be redshifted into near-infrared regime.
We have found 30 mntBALs on the redshift ranges.

\section{Our results}
We discovered H$\alpha$ absorption line in SDSS~0839+3805 \citep{Aok06}
and 
Balmer-series absorption lines from H$\alpha$ and H9 in SDSS~1723+5553
(Fig. 1).
Presence of the non-stellar Balmer absorption lines is known to date only in the Seyfert galaxy NGC 4151, 
thus our discovery is the first cases for quasars.
Recently \citet{Hal06} reports Balmer absorption lines in another FeLoBAL,
SDSS~J1259+1213.
That quasar is also a possible mntBAL.
The redshifts are determined to be 
2.3179 and 2.1081 for SDSS~0839+3805 and SDSS~1723+5553, respectively,
by H$\alpha$ emission line.
The Balmer absorption lines are blueshifted from the emission line
by 520 and 5370 km s$^{-1}$, and
the column density of neutral hydrogen is estimated to be $10^{18}$
and $5 \times 10^{19}$ cm$^{-2}$ for SDSS~0839+3805 and SDSS~1723+5553,
respectively.
\par
At least four active galactic nuclei (NGC~4151, SDSS~0839+3805, SDSS~J1723+5553, and SDSS~J1259+1213) have known to date show non-stellar Balmer absorption lines.
Although they are much different in their luminosity, 
they share several characteristics.
All four objects have Fe {\small II} absorption lines.
NGC~4151 is not classified as FeLoBALs, however, its spectrum 
at the low luminosity phase shows many Fe {\small II} absorption lines 
\citep{Kra01}.
It is remarkably similar to SDSS~0839+3805 \citep{Aok06}.
Strong [O {\small III}] emission line exists in all four objects.
This attribute makes contrast to the previous observations of very
weak or non-detection [O {\small III}] in
LoBALs/FeLoBALs \citep{BM92, YW03}.
The FeLoBALs with Balmer absorption lines do not have strong Fe {\small II} optical emission lines.
These facts suggest that the existence of such a population among BAL quasars,
and they have unique structures at their nuclei 
or unique evolutionary phase.
\begin{figure}[!ht]
\plotone{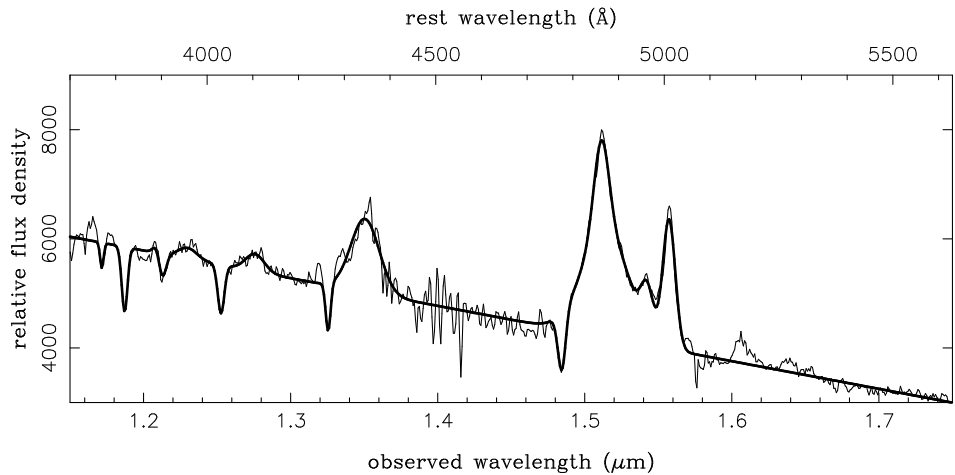}
\caption{Rest-optical spectrum of SDSS~J1723+5553 in $JH$-band.
Ordinate is a relative flux density in units of erg s$^{-1}$ cm$^{-2}$
$\mu$m$^{-1}$
and abscissa is an observed wavelength in vacuum in micron.
The rest wavelength is given along the top axis.
The best fit is plotted as a thick solid line.}
\end{figure}



\end{document}